\documentclass[runningheads]{llncs}
\usepackage[T1]{fontenc}
\usepackage{colortbl}
\usepackage{array,multirow,graphicx}
\usepackage{capt-of}
\usepackage{color}
\usepackage{algorithm}
\usepackage[noend]{algpseudocode}
\usepackage{hyperref}
\usepackage{tikz}
\usepackage{comment,nicefrac}
\usepackage{graphicx}
\usepackage{capt-of}
\usepackage{bbm}
\usepackage{amsmath,amssymb}
\usepackage{booktabs}
\usepackage{epigraph,booktabs}
\usepackage{rotating}
\usepackage{tabularx}
\usepackage{adjustbox}
\begin{document}
\title{Annotator Consensus Prediction for Medical Image Segmentation with Diffusion Models}
\titlerunning{Annotator Consensus Prediction}
%

\author{Tomer Amit \and
Shmuel Shichrur \and
Tal Shaharbany \and Lior Wolf}
\authorrunning{Amit et al.}

\institute{Tel-Aviv University\\\email{\{tomeramit1,shmuels1,shaharabany,wolf\}@mail.tau.ac.il}}


%
%
\maketitle              
\begin{abstract}
A major challenge in the segmentation of medical images is the large inter- and intra-observer variability in annotations provided by multiple experts. To address this challenge, we propose a novel method for multi-expert prediction using diffusion models. Our method leverages the diffusion-based approach to incorporate information from multiple annotations and fuse it into a unified segmentation map that reflects the consensus of multiple experts. We evaluate the performance of our method on several datasets of medical segmentation annotated by multiple experts and compare it with the state-of-the-art methods. Our results demonstrate the effectiveness and robustness of the proposed method. Our code is publicly available at \url{https://github.com/tomeramit/Annotator-Consensus-Prediction}

\end{abstract}
\section{Introduction}

Medical image segmentation is a challenging task that requires accurate delineation of structures and regions of interest in complex and noisy images. Multiple expert annotators are often employed to address this challenge, to provide binary segmentation annotations for the same image. However, due to differences in experience, expertise, and subjective judgments, annotations can vary significantly, leading to inter- and intra-observer variability. In addition, manual annotation is a time-consuming and costly process, which limits the scalability and applicability of segmentation methods.

To overcome these limitations, automated methods for multi-annotator prediction have been proposed, which aim to fuse the annotations from multiple annotators and generate an accurate and consistent segmentation result. Existing approaches for multi-annotator prediction include majority voting~\cite{guan2018said}, label fusion~\cite{chen2019automatic}, and label sampling~\cite{jensen2019improving}. 

In recent years, diffusion models have emerged as a promising approach for image segmentation, for example by using learned semantic features~\cite{baranchuk2021label}. By modeling the diffusion of image intensity values over the iterations, diffusion models capture the underlying structure and texture of the images and can separate regions of interest from the background. Moreover, diffusion models can handle noise and image artifacts, and adapt to different image modalities and resolutions.

In this work, we propose a novel method for multi-annotator prediction, using diffusion models for medical binary segmentation. The goal of multi-annotator prediction is to fuse multiple annotations of the same image from different annotators and obtain a more accurate and reliable segmentation result. In practice, we leverage the diffusion-based approach to create one map for each level of consensus. To obtain the final prediction, we average the obtained maps and obtain one soft map. 

We evaluate the performance of the proposed method on a dataset of medical images annotated by multiple annotators. Our results demonstrate the effectiveness and robustness of the proposed method in handling inter- and intra-observer variability and achieving higher segmentation accuracy than the state-of-the-art methods. The proposed method could improve the efficiency and quality of medical image segmentation and facilitate the clinical decision-making process.

\section{Related work}
\smallskip\noindent{\bf Multi-annotator strategies\quad} Research attention has recently been directed towards the issues of multi-annotator labels~\cite{guan2018said,jensen2019improving}. During training, Jensen et al.~\cite{jensen2019improving} randomly sampled different labels per image. This method produced a more calibrated model. Guan et al.~\cite{guan2018said} predicted the gradings of each annotator individually and acquired the corresponding weights for the final prediction. Kohl et al.~\cite{kohl2018probabilistic} used the same sampling strategy to train a probabilistic model, based on a U-Net combined with a conditional variational autoencoder. Another recent probabilistic approach~\cite{rahman2023ambiguous} combines a diffusion model with KL divergence to capture the variability between the different annotators. In our work, we use consensus maps as the ground truth and compare them to other strategies.

\smallskip\noindent{\bf Diffusion Probabilistic Models (DPM)~\cite{sohl2015deep}\quad} are a class of generative models based on a Markov chain, which can transform a simple distribution (e.g. Gaussian) to data sampled from a complex distribution. Diffusion models are capable of generating high-quality images that can compete with and even outperform the latest GAN methods~\cite{sohl2015deep,ho2020denoising,nichol2021improved,dhariwal2021diffusion}. A variational framework for the likelihood estimation of diffusion models was introduced by Huang et al.~\cite{huang2021variational}. Subsequently, Kingma et al.~\cite{kingma2021variational} proposed a Variational Diffusion Model that produces state-of-the-art results in likelihood estimation for image density. 


\smallskip\noindent{\bf Conditional Diffusion Probabilistic Models\quad} In our work, we use diffusion models to solve the image segmentation problem as conditional generation, given the image. Conditional generation with diffusion models includes methods for class-conditioned generation, which is obtained by adding a class embedding to the timestep embedding~\cite{nichol2021improved}. In \cite{choi2021ilvr}, a method for guiding the generative process in DDPM is present. This method allows the generation of images based on a given reference image without any additional learning. In the domain of super-resolution, the lower-resolution image is upsampled and then concatenated, channelwise, to the generated image at each iteration~\cite{saharia2021image,ho2022cascaded}. A similar approach passes the low-resolution images through a convolutional block~\cite{li2022srdiff} prior to the concatenation. 


A previous study directly applied a diffusion model to generate a segmentation mask based on a conditioned input image~\cite{amit2021segdiff}. Baranchuk et al.~\cite{baranchuk2021label} extract features from a pretrained diffusion model for training a segmentation network, while our diffusion model generates the output mask. Compared to the diffusion-based image segmentation method of Wolleb et al.~\cite{wolleb2021diffusion}, our architecture differs in two main aspects: (i) the concatenation method of the condition signal, and (ii) an encoder that processes the conditioning signal. We also use a lower value of T, which reduces the running time.

\begin{figure*}[t]
    \centering
    \includegraphics[width=0.99\linewidth]{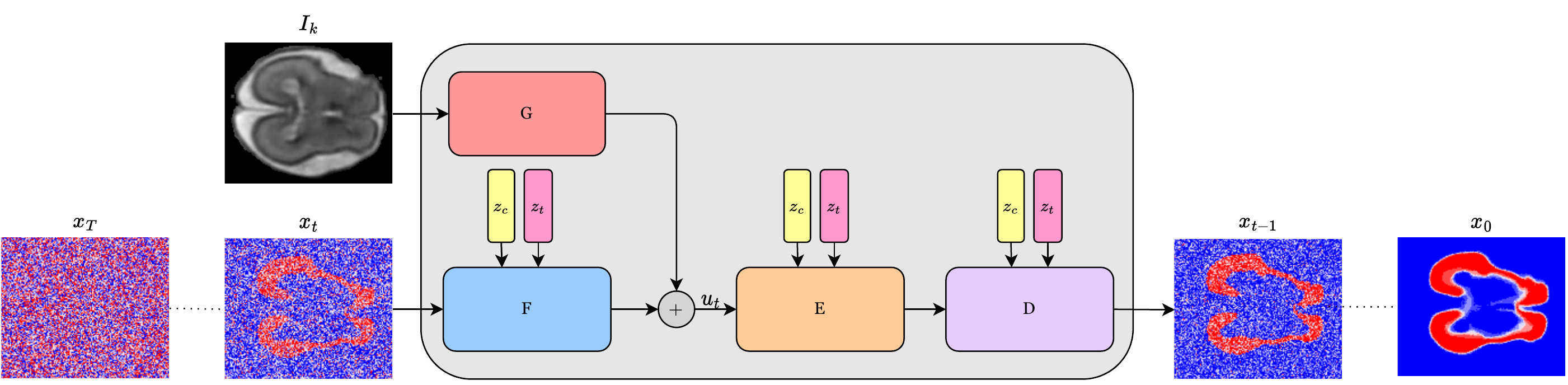}
    \caption{The figure below illustrates our proposed method for multi-annotator segmentation. The input $I_k$ image with the noisy segmentation map $x_t$ is passed through our network iteratively $T$ times in order to obtain an output segmentation map $x_0$. Each network receives the consensus level $c$ as an embedding $z_c$ as well as the time step data.}
    \label{fig:arch}
\end{figure*}

\section{Method}
Our approach for binary segmentation with multi-annotators employs a diffusion model that is conditioned on the input image $I\in  R^{W \times H}$, the step estimation $t$, and the consensus index $c$. The diffusion model updates its current estimate $x_t$ iteratively, using the step estimation function $\epsilon_\theta$. See Fig.~\ref{fig:arch} for an illustration. 

Given a set of C annotations $\{A_k^i\}_{i=1}^C$ associated with input sample $I_k$, we define the ground truth consensus map at level $c$ to be
\begin{equation}
M_{k}^c[x,y] = 
\begin{cases}
1 & \sum_{i=1}^C A^i_k[x,y] \geq c,\\
0 & \text{otherwise},
\end{cases}
\end{equation}

During training, our algorithm iteratively samples a random level of the consensus $c\sim U[1,2,...,C]$ and an input-output pair $(I_k, M_{k}^c)$. The iteration number $1\leq t\leq T$ is sampled from a uniform distribution and $X_T$ is sampled from a normal distribution. 

We then compute $x_t$ from $X_T$, $M_{k}^c$ and $t$ according to:

\begin{equation}
    x_t 
    = 
    \sqrt{\Bar{\alpha}_t}M_{k}^c + \sqrt{(1-\Bar{\alpha}_t)} X_T, X_T \thicksim N(0,I_{n\times n}).
\label{eq:xt_reparametrization}
\end{equation}
where $\Bar{\alpha}$ is a constant that defines the schedule of added noise.

The current step index $t$, and the consensus index $c$ are integers that are translated to $z_t\in R^d$ and $z_c\in R^d$, respectively with a pair of lookup tables. 
The embeddings are passed to the different networks $F$, $D$ and $E$.

In the next step, our algorithm encodes the input signal $x_t$ with network $F$ and encodes the condition image $I_k$ with network $G$. We compute the conditioned signal $u_t = F(x_t, z_c, z_t)+G(I_k)$,
and apply it to the networks $E$ and $D$, where the output is the estimation of $x_{t-1}$. 

\begin{equation}
\label{eq:newepsilon}
    \epsilon_\theta(x_t, I_k, z_t, z_c) = D(E(F(x_t, z_t, z_c)+G(I_k), z_t, z_c), z_t, z_c)\,.
\end{equation}
 
The loss function being minimized is:
\begin{equation}
    E_{x_0,\epsilon,x_e,t,c}[||\epsilon-\epsilon_\theta(x_t, I_k, z_t, z_c)||^2].
\label{eq:conditional_loss_term}
\end{equation}



The training procedure is depicted in Alg.~\ref{alg:Training}. The total number of diffusion steps $T$ is set by the user, and C is the number of different annotators in the dataset. Our model is trained using binary consensus maps ($M^c_{k}$) as the ground truth, where $k$ is the sample id, and $c$ is the consensus index.

The inference process is described in Alg.~\ref{alg:Inference}. We sample our model for each consensus index, and then calculate the mean of all results to obtain our target, which is a soft-label map representing the annotator agreement. Mathematically, if the consensus maps are perfect, this is equivalent to assigning each image location with the fraction of annotations that consider this location to be part of the mask (if $c$ annotators mark a pixel, it would appear in levels $1..c$). In Section~\ref{sec:ablation}, we compare our method with other variants and show that estimating the fraction map directly, using an identical diffusion model, is far inferior to estimating each consensus level separately and then averaging.

\begin{table}[t]
\begin{minipage}[t]{0.5\textwidth}
\begin{algorithm}[H]
    \caption{Training Algorithm}
    \begin{algorithmic}
        \State \textbf{Input} $T$, $D = \{(I_k,M^1_{k},...,M^C_{k})\}^K_k$
        \Repeat
            \State sample $c \thicksim \{1,...,C\}$
            \State sample $(I_k,M^c_{k}) \thicksim D'$
            \State sample $\epsilon \thicksim N(\mathbf{0},\mathbf{I_{n\times n}})$ 
            \State sample $t \thicksim $ (\{1,...,T\})
            \State $z_c = LUT_c(c)$
            \State $z_t = LUT_t(t)$
            \State $\beta_t=\frac{10^{-4}(T-t) + 2*10^{-2}(t-1)}{T-1}$
            \State $\alpha_t  = 1-\beta_t$
            \State $\Bar{\alpha}_t = \prod^t_{s=0}\alpha_s$
            \State $x_t=\sqrt{\Bar{\alpha_t}}M^c_k + \sqrt{1-\Bar{\alpha_t}}\epsilon$
            \State $\nabla_\theta|| \epsilon -\epsilon_\theta (x_t,I_k,z_t,z_c)||$
        \Until{convergence}
    \end{algorithmic}

\label{alg:Training}
\end{algorithm}
\end{minipage}
\begin{minipage}[t]{0.50\textwidth}
\begin{algorithm}[H]
    \caption{Inference Algorithm}
    \begin{algorithmic}
        \State \textbf{Input} $T$, $I$
        \For{$c=1,...,C$}
        \State sample $x_{T_c} \thicksim N(\mathbf{0},\mathbf{I_{n\times n}})$
            \For{$t=T,T-1,...,1$}
                \State sample $z \thicksim N(\mathbf{0},\mathbf{I_{n\times n}})$
                \State $z_c = LUT_c(c)$, $z_t = LUT_t(t)$
                \State $\beta_t=\frac{10^{-4}(T-t) + 2*10^{-2}(t-1)}{T-1}$
                \State $\alpha_t  = 1-\beta_t$.  $\Bar{\alpha}_t = \prod^t_{s=0}\alpha_s$
                \State $\tilde{\beta}_t = \frac{1-\Bar{\alpha}_{t-1}}{1-\Bar{\alpha}_t}\beta_t$
                \State $\epsilon'_t = \frac{1-\alpha_t}{\sqrt{1-\Bar{\alpha}_t}}\epsilon_\theta(\Bar{x}_t,I,z_t,z_c)$
                \State $\Bar{x}_{{t-1}_c} = {\alpha_t}^{-\scriptscriptstyle\frac{1}{2}}(x_t - \epsilon'_t)$
                
                \State $x_{{t-1}_c} = \Bar{x}_{{t-1}_c} + \mathbbm{1}_{[t > 1]}{\tilde{\beta}_t}^{\scriptscriptstyle\frac{1}{2}} z$
                
            \EndFor

        \EndFor
        \State \Return $(\sum_{i=1}^C x_{0_i}) / C$
    \end{algorithmic}
\label{alg:Inference}
\end{algorithm}
\end{minipage}
\end{table}

\begin{figure*}[t]
\centering
\begin{tabular}{@{}c@{~}c@{~}c@{~}c@{~}c@{~}c@{~}c@{~}c@{}}

        \rotatebox[origin=l]{90}{Kidney} & \includegraphics[width=0.15223235312\textwidth]{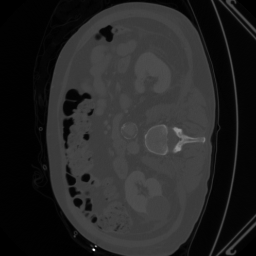} & \includegraphics[width=0.15223235312\textwidth]{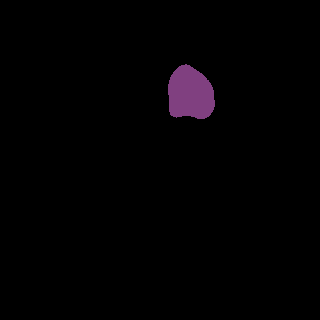} & \includegraphics[width=0.15223235312\textwidth]{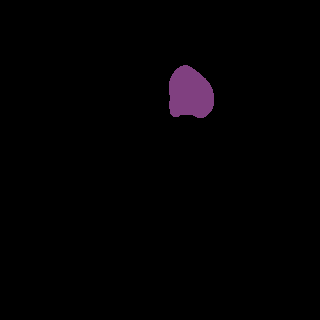} & \includegraphics[width=0.15223235312\textwidth]{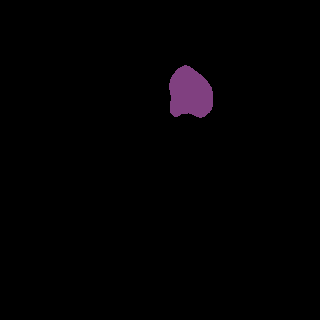} & \includegraphics[width=0.15223235312\textwidth]{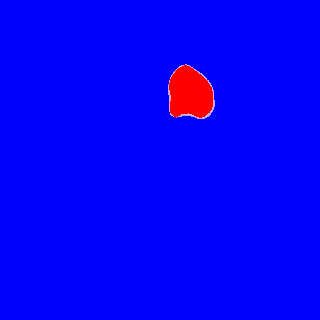} &  \includegraphics[width=0.15223235312\textwidth]{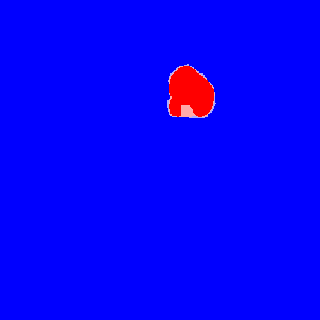}\\

        \rotatebox[origin=l]{90}{Brain} & \includegraphics[width=0.15223235312\textwidth]{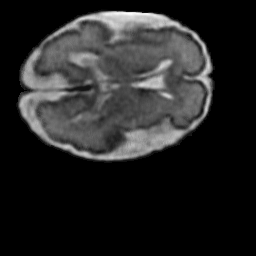} & \includegraphics[width=0.15223235312\textwidth]{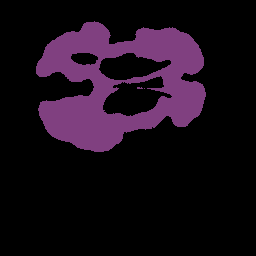} & \includegraphics[width=0.15223235312\textwidth]{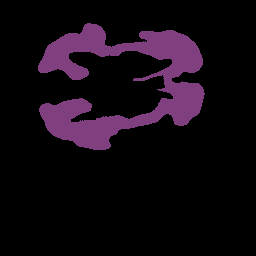} & \includegraphics[width=0.15223235312\textwidth]{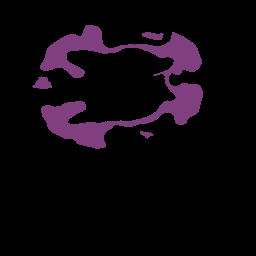} & \includegraphics[width=0.15223235312\textwidth]{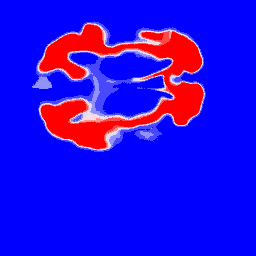} &  \includegraphics[width=0.15223235312\textwidth]{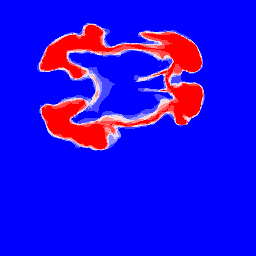}\\
        
        \rotatebox[origin=l]{90}{Tumor} & \includegraphics[width=0.15223235312\textwidth]{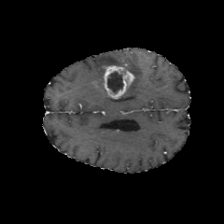} & \includegraphics[width=0.15223235312\textwidth]{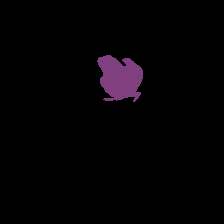} & \includegraphics[width=0.15223235312\textwidth]{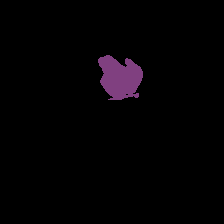} & \includegraphics[width=0.15223235312\textwidth]{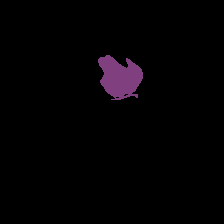} & \includegraphics[width=0.15223235312\textwidth]{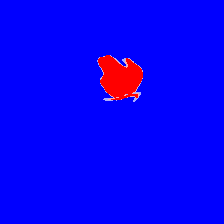} &  \includegraphics[width=0.15223235312\textwidth]{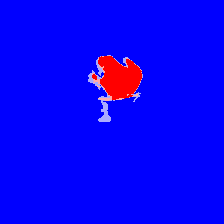}\\
        
        \rotatebox[origin=l]{90}{Prostate 1} & \includegraphics[width=0.15223235312\textwidth]{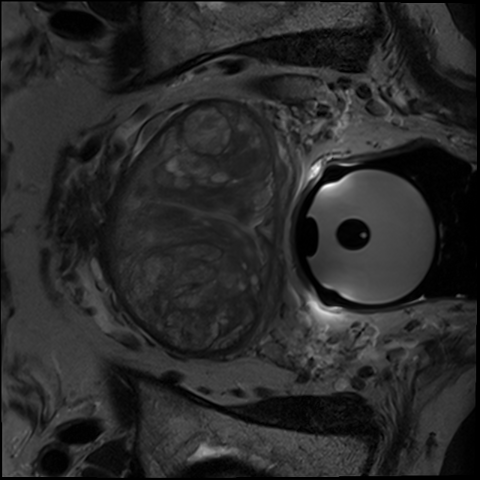} & \includegraphics[width=0.15223235312\textwidth]{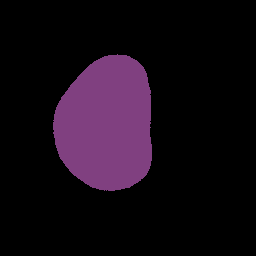} & \includegraphics[width=0.15223235312\textwidth]{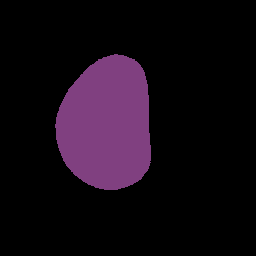} & \includegraphics[width=0.15223235312\textwidth]{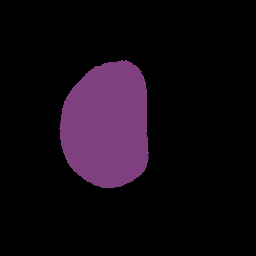} & \includegraphics[width=0.15223235312\textwidth]{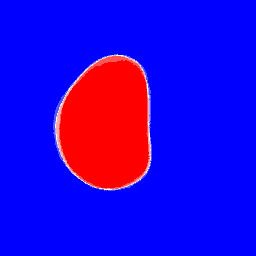} &  \includegraphics[width=0.15223235312\textwidth]{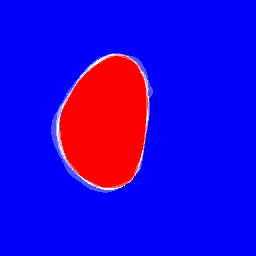}\\
        \rotatebox[origin=l]{90}{Prostate 2} & 
        \includegraphics[width=0.15223235312\textwidth]{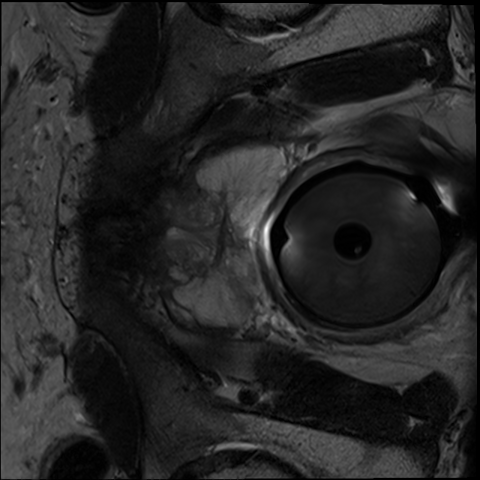} & \includegraphics[width=0.15223235312\textwidth]{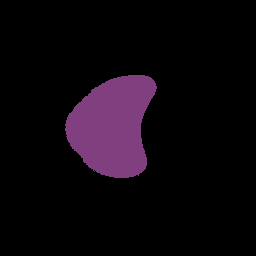} & \includegraphics[width=0.15223235312\textwidth]{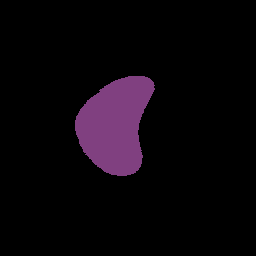} & \includegraphics[width=0.15223235312\textwidth]{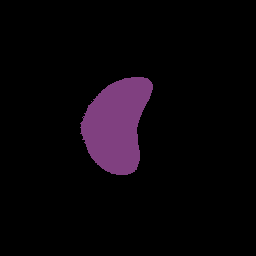} & \includegraphics[width=0.15223235312\textwidth]{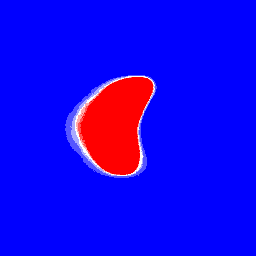} &  \includegraphics[width=0.15223235312\textwidth]{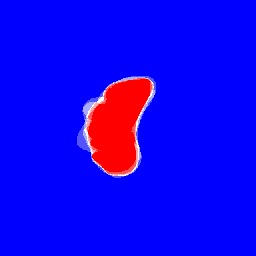}\\

       \cmidrule(lr){1-1} \cmidrule(lr){2-2} \cmidrule(lr){3-5}\cmidrule(lr){6-6}\cmidrule(lr){7-7}
        (a) & (b) & \multicolumn{3}{c}{(c)} & (d) & (e)\\        

\end{tabular}
\caption{Multiple segmentation results on all datasets of the QUBIQ benchmark. (a) dataset, (b) input image, (c) a subset of the obtained consensus maps for multiple runs with different consensus index on the same input, (d) average result, visualized by the 'bwr' color scale between 0 (blue) and 1 (red), and (e) ground truth.}
\label{fig:variabiltiy}
\end{figure*}

\smallskip\noindent{\bf Employing multiple generations\quad}Since calculating $x_{t-1}$ during inference includes the addition of 
$\mathbbm{1}_{[t > 1]}{\tilde{\beta}_t}^{\scriptscriptstyle\frac{1}{2}} z$ where $z$ is from a standard distribution, there is significant variability between different runs of the inference method on the same inputs, see Fig.~\ref{fig:variabiltiy}(b). 
 
In order to exploit this phenomenon, we run the inference algorithm multiple times, then average the results. 
This way, we stabilize the results of segmentation and improve performance, as demonstrated in Fig.~\ref{fig:variabiltiy}(c). We use twenty-five generated instances in all experiments. In the ablation study, we quantify the gain of this averaging procedure.



\smallskip\noindent{\bf Architecture\quad} In this architecture, the U-Net's decoder $D$ is conventional and its encoder is broken down into three networks: $E$, $F$, and $G$. The last encodes the input image, while $F$ encodes the segmentation map of the current step $x_t$. The two processed inputs have the same spatial dimensionality and number of channels. Based on the success of residual connections~\cite{he2016deep}, we sum these signals $F(x_t, z_t, z_c)+G(I)$. This sum then passes to the rest of the U-Net encoder $E$.

The input image encoder $G$ is built from Residual in Residual Dense Blocks~\cite{wang2018esrgan} (RRDBs), which combine multi-level residual connections without batch normalization layers. $G$ has an input 2D-convolutional layer, an RRDB with a residual connection around it, followed by another 2D-convolutional layer, leaky RELU activation and a final 2D-convolutional output layer. $F$ is a 2D-convolutional layer with a single-channel input and an output of $L$ channels.

The encoder-decoder part of $\epsilon_\theta$, i.e., $D$ and $E$, is based on U-Net, similarly to~\cite{nichol2021improved}. Each level is composed of residual blocks, and at resolution 16x16 and 8x8 each residual block is followed by an attention layer. The bottleneck contains two residual blocks with an attention layer in between. Each attention layer contains multiple attention heads.

The residual block is composed of two convolutional blocks, where each convolutional block contains group-norm, SiLU activation, and a 2D-convolutional layer. The residual block receives the time embedding through a linear layer, SiLU activation, and another linear layer. The result is then added to the output of the first 2D-convolutional block. Additionally, the residual block has a residual connection that passes all its content.

On the encoder side (network $E$), there is a downsample block after the residual blocks of the same depth, which is a 2D-convolutional layer with a stride of two. On the decoder side (network $D$), there is an upsample block after the residual blocks of the same depth, which is composed of the nearest interpolation that doubles the spatial size, followed by a 2D-convolutional layer. Each layer in the encoder has a skip connection to the decoder side.

\section{Experiments}
We conducted a series of experiments to evaluate the performance of our proposed method for multi-annotator prediction. Our experiments were carried out on datasets of the QUBIQ benchmark\footnote{Quantification of Uncertainties in Biomedical Image Quantification Challenge in MICCAI20'- \href{https://qubiq.grand-challenge.org/}{link}}. We compared the performance of our proposed method with several state-of-the-art methods.
 
\label{sec:experiments}
\smallskip\noindent{\bf Datasets\quad} The Quantification of Uncertainties in Biomedical Image Quantification Challenge (QUBIQ), is a recently available challenge dataset specifically for the evaluation of inter-rater variability. QUBIQ comprises four different segmentation datasets with CT and MRI modalities, including brain growth (one task, MRI, seven raters, 34 cases for training and 5 cases for testing), brain tumor (one task, MRI, three raters, 28 cases for training and 4 cases for testing), prostate (two subtasks, MRI, six raters, 33 cases for training and 15 cases for testing), and kidney (one task, CT, three raters, 20 cases for training and 4 cases for testing).

Following~\cite{ji2021learning}, the evaluation is performed using the soft Dice coefficient with five threshold levels, set as (0.1, 0.3, 0.5, 0.7, 0.9).

\smallskip\noindent{\bf Implementation details\quad}
The number of diffusion steps in previous works was 1000~\cite{ho2020denoising} and even 4000~\cite{nichol2021improved}. The literature suggests that more is better~\cite{san2021noise}. In our experiments, we employ 100 diffusion steps, to reduce inference time.

The AdamW~\cite{loshchilov2017decoupled} optimizer is used in all our experiments. Based on the intuition that the more RRDB blocks, the better the results, we used as many blocks as we could fit on the GPU without overly reducing batch size.

Following ~\cite{ji2021learning}, for all datasets of the QUBIQ benchmark the input image resolution, as well as the test image resolution, was $256 \times 256$. The experiments were performed with a batch size of four images and eight RRDB blocks. The network depth was seven, and the number of channels in each depth was $[L, L, L, 2L, 2L, 4L, 4L]$, with $L=128$. The augmentations used were: random scaling by a factor sampled uniformly in the range $[0.9, 1.1]$, a rotation between 0 and 15 degrees, translation between $[0, 0.1]$ in both axes, and horizontal and vertical flips, each applied with a probability of 0.5.


\smallskip\noindent{\bf Results\quad} We compare our method with 
FCN~\cite{long2015fully}, MCD~\cite{gal2016dropout}, FPM~\cite{zhou2017fixed}, DAF~\cite{wang2018deep}, MV-UNet~\cite{ji2021learning},
LS-UNet~\cite{jensen2019improving},
MH-UNet~\cite{guan2018said}, and
MRNet~\cite{ji2021learning}.

\begin{figure}[t]
  \centering
  \begin{minipage}[c]{0.47\linewidth}
\centering
    \includegraphics[width=0.98\linewidth]{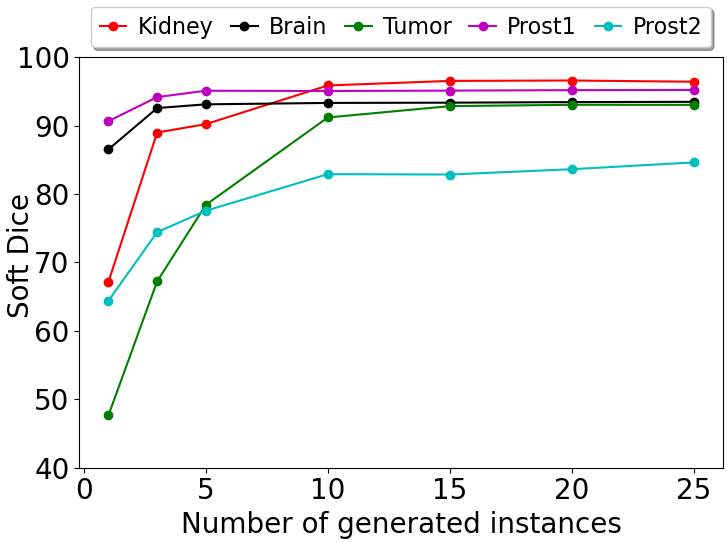}
\caption{Soft Dice vs. \#generated images.}
\label{fig:generated_images}
  \end{minipage}%
\hfill
\begin{minipage}[c]{0.51\linewidth}
\centering
\begin{adjustbox}{width=0.98\linewidth}
\begin{tabular}{@{}l@{~}c@{~}c@{~}c@{~}c@{~}c@{}}
\hline
Method & Kidney & Brain & Tumor & Prost1 & Prost2 \\
\hline
FCN & 70.03 & 80.99 & 83.12 & 84.55 & 67.81 \\
MCD & 72.93 & 82.91 & 86.17 & 86.40 & 70.95 \\
FPM & 72.17 & - & - & - & - \\
DAF & - & - & - & 85.98 & 72.87 \\
MV-UNet & 70.65 & 81.77 & 84.03 & 85.18 & 68.39 \\
LS-UNet & 72.31 & 82.79 & 85.85 & 86.23 & 69.05 \\
MH-UNet & 73.44 & 83.54 & 86.74 & 87.03 & 75.61 \\
MRNet & 74.97 & 84.31 & 88.40 & 87.27 & 76.01 \\
AMIS & 68.53 & 74.09 & 92.95 & 91.64 & 21.91 \\
DMISE & 74.50 & 92.80 & 87.80 & 94.70 & 80.20 \\
Ours & \textbf{96.58} & \textbf{93.81} & \textbf{93.16} & \textbf{95.21} & \textbf{84.62} \\
\hline
\end{tabular}
\end{adjustbox}
\captionof{table}{QUBIQ soft Dice results.}
  \label{tab:QUBIQ}
\end{minipage}
\end{figure}

We also compare with models that we train ourselves, using public code AMIS~\cite{rahman2023ambiguous}, and DMISE~\cite{wolleb2021diffusion}. The first is trained in a scenario where each annotator is a different sample (``No annotator'' variant of our ablation results below), and the second is trained on the consensus setting, similar to our method. As can be seen in Tab.~\ref{tab:QUBIQ}, our method outperforms all other methods across all datasets of QUBIQ benchmark.  



\smallskip\noindent{\bf Ablation Study\quad} 
\label{sec:ablation}
We evaluate alternative training variants as an ablation study in Tab~\ref{tab:ablation}. The ``Annotator'' variant, in which our model learns to produce each annotator binary segmentation map and then averages all the results to obtain the required soft-label map, achieves lower scores compared to the ``Consensus'' variant, which is our full method. The ``No annotator'' variant, where images were paired with random annotators without utilizing the annotator IDs, achieves a slightly lower average score compared to the ``Annotator'' variant. We also note that our ``No annotator'' variant outperforms the analog AMIS model in four out of five datasets, indicating that our architecture is somewhat preferable. In a third variant, our model learns to predict the soft-label map that denotes the fraction of annotators that mark each image location directly. Since this results in fewer generated images, we generate $C$ times as many images per test sample. The score of this variant is also much lower than that of our method. 

\begin{figure*}[t]
\centering

\begin{tabular}{c@{}c@{~}c@{~}c@{~}c@{~}c@{~}c@{~}c@{}}
\rotatebox[origin=l]{90}{Kidney} 
        &\includegraphics[width=0.15223235312\textwidth]{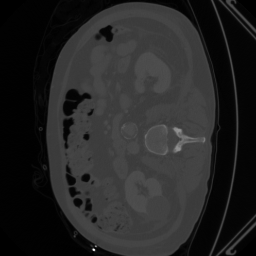} & 
        \includegraphics[width=0.15223235312\textwidth]{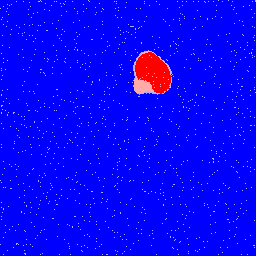} & \includegraphics[width=0.15223235312\textwidth]{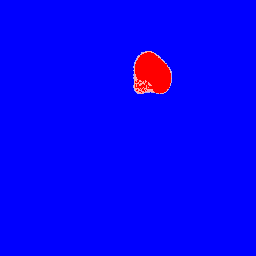} & \includegraphics[width=0.15223235312\textwidth]{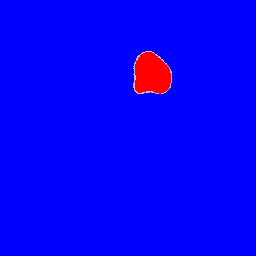} & \includegraphics[width=0.15223235312\textwidth]{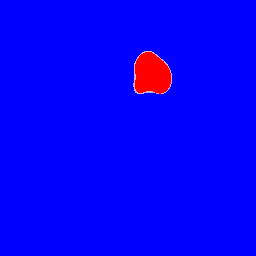} & \includegraphics[width=0.15223235312\textwidth]{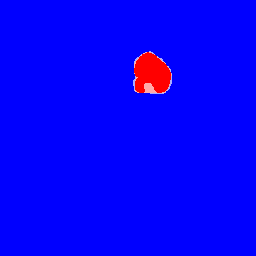}\\

        \rotatebox[origin=l]{90}{Brain} &\includegraphics[width=0.15223235312\textwidth] {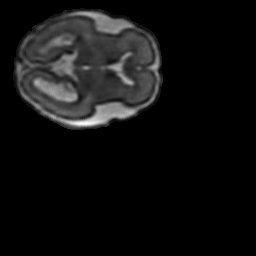} &
        \includegraphics[width=0.15223235312\textwidth]{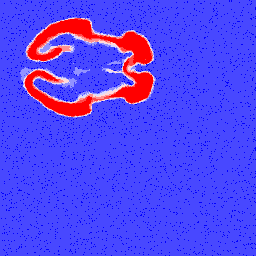} & \includegraphics[width=0.15223235312\textwidth]{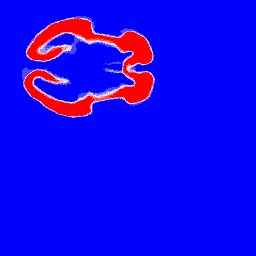} & \includegraphics[width=0.15223235312\textwidth]{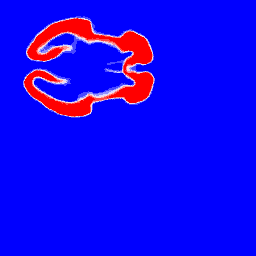} & \includegraphics[width=0.15223235312\textwidth]{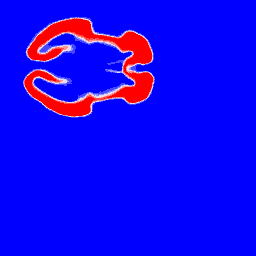} & \includegraphics[width=0.15223235312\textwidth]{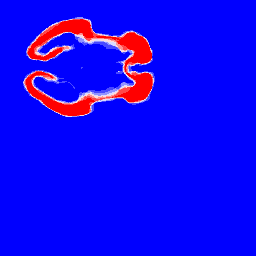}\\

        \rotatebox[origin=l]{90}{Tumor} 
        &\includegraphics[width=0.15223235312\textwidth]{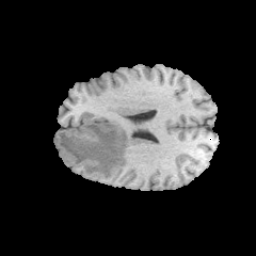} & 
        \includegraphics[width=0.15223235312\textwidth]{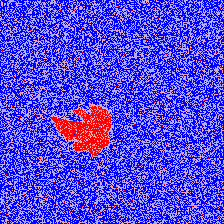} & \includegraphics[width=0.15223235312\textwidth]{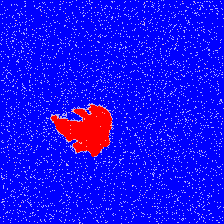} & \includegraphics[width=0.15223235312\textwidth]{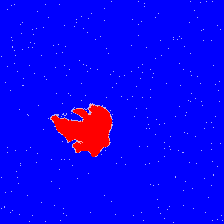} & \includegraphics[width=0.15223235312\textwidth]{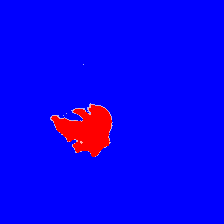} & \includegraphics[width=0.15223235312\textwidth]{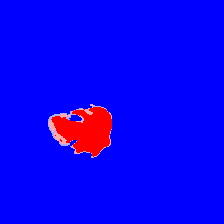}\\

\rotatebox[origin=l]{90}{Prostate 1}         
        &\includegraphics[width=0.15223235312\textwidth]{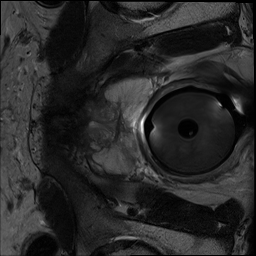} & 
        \includegraphics[width=0.15223235312\textwidth]{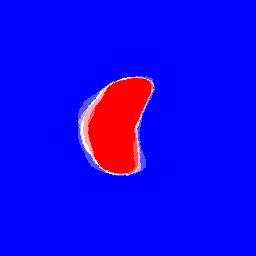} & \includegraphics[width=0.15223235312\textwidth]{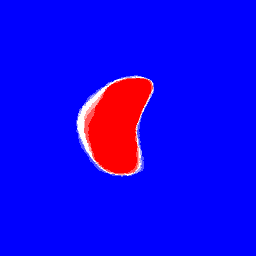} & \includegraphics[width=0.15223235312\textwidth]{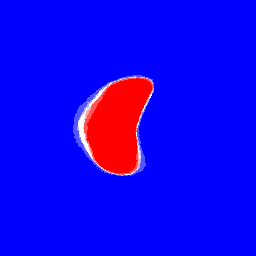} & \includegraphics[width=0.15223235312\textwidth]{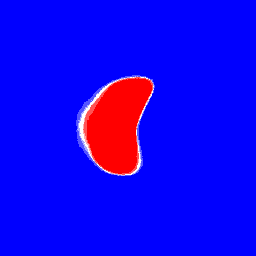} & \includegraphics[width=0.15223235312\textwidth]{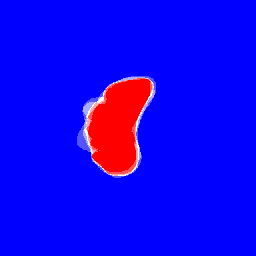}\\
\rotatebox[origin=l]{90}{Prostate 2} 
        &\includegraphics[width=0.15223235312\textwidth]{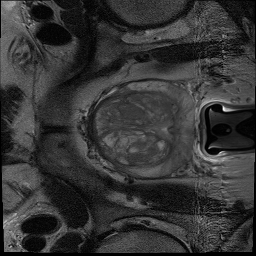} & 
        \includegraphics[width=0.15223235312\textwidth]{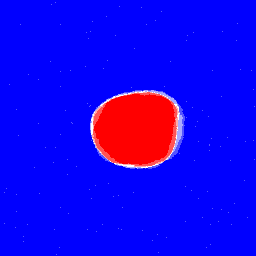} & \includegraphics[width=0.15223235312\textwidth]{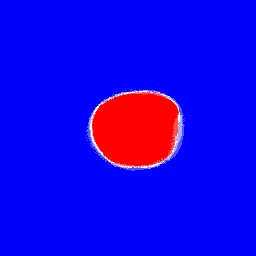} & \includegraphics[width=0.15223235312\textwidth]{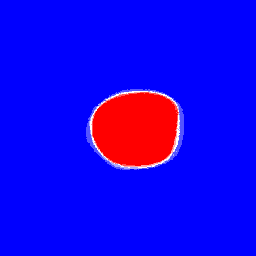} & \includegraphics[width=0.15223235312\textwidth]{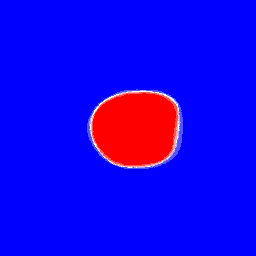} & \includegraphics[width=0.15223235312\textwidth]{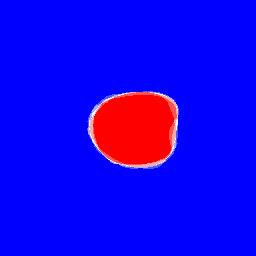}\\
        
       \cmidrule(lr){1-1} \cmidrule(lr){2-2} \cmidrule(lr){3-6} \cmidrule(lr){7-7}
        (a) & (b) & \multicolumn{4}{c}{(c)} & (d)\\        

\end{tabular}
\caption{Multiple segmentation results per number of generated images. (a) dataset, (b) input image, (c) results for 1, 5, 10, 25 generated images, and (d) ground truth.}
\label{fig:example_number_generated_images}
\end{figure*}

Next, we study the effect of the number of generated images on performance. The results can be seen in Fig.~\ref{fig:generated_images}. In general, increasing the number of generated instances tends to improve performance. However, the number of runs required to reach optimal performance varies between classes. For example, for the Brain and the Prostate 1 datasets, optimal performance is achieved using 5 generated images, while on Prostate 2 the optimal performance is achieved using 25 generated images.  Fig.~\ref{fig:example_number_generated_images} depicts samples from multiple datasets and presents the progression as the number of generated images increases. As can be seen, as the number of generated images increases, the outcome becomes more and more similar to the target segmentation.

   \begin{table}[t]
  \label{tab:training_variants}
\centering
\begin{small}
\begin{tabular}{lccccc}
\hline
Method & Kidney & Brain & Tumor & Prostate 1 & Prostate 2 \\
\hline

Annotator & 96.13 & 89.88 & 92.51 & 93.89 & 76.89 \\
No annotator & 94.46 & 89.78 & 91.78 & 92.58 & 78.61 \\
Soft-label & 65.41 & 79.56 & 75.60 & 73.23 & 65.24 \\
Consensus (our method) & \textbf{96.58} & \textbf{93.81} & \textbf{93.16} & \textbf{95.21} & \textbf{84.62} \\
\hline
\end{tabular}
\end{small}
\vspace{0.3cm}
\captionof{table}{Ablation study showing soft Dice results for various alternative methods of training similar diffusion models.}
\label{tab:ablation}
 \end{table}

   \begin{table}[t]
 \centering
\begin{tabular}{lc}
\toprule
Dataset & Mean score between pairs\\
\midrule
Kidney & 94.95 \\
Brain & 85.74 \\
Tumor & 90.65   \\
Prostate 1 & 94.64 \\
Prostate 2 & 89.91 \\
\bottomrule
\end{tabular}
\vspace{0.3cm}
\caption{Pairwise Dice scores per dataset.}
\label{tab:pairwise_dice}
 \end{table}

\section{Discussion}

In order to investigate the relationship between the annotator agreement and the performance of our model, we conducted an analysis by calculating the average Dice score between each pair of annotators across the entire dataset. The results of this pairwise Dice analysis can be found in Tab~\ref{tab:pairwise_dice}, where higher mean-scores indicate a greater consensus among the annotators.

We observed that our proposed method demonstrated improved performance on datasets with higher agreement among annotators, specifically the kidney and prostate 1 datasets. Conversely, the performance of the other methods significantly deteriorated on the kidney dataset, leading to a lower correlation between the Dice score and the overall performance.

Additionally, we examined the relationship between the number of annotators and the performance of our model. Surprisingly, we found no significant correlation between the number of annotators and the performance of our model.

\section{Conclusions}
Shifting the level of consensus required to mark a region from very high to as low as one annotator, can be seen as creating a dynamic shift from a very conservative segmentation mask to a very liberal one. As it turns out, this dynamic is well-captured by diffusion models, which can be readily conditioned on the level of consensus. Another interesting observation that we make is that the mean (over the consensus level) of the obtained consensus masks is an effective soft mask. Applying these two elements together, we obtain state-of-the-art results on multiple binary segmentation tasks. 
%
%
%
\bibliographystyle{splncs04}
\bibliography{mybibliography}

\begin{thebibliography}{10}
\providecommand{\url}[1]{\texttt{#1}}
\providecommand{\urlprefix}{URL }
\providecommand{\doi}[1]{https://doi.org/#1}

\bibitem{amit2021segdiff}
Amit, T., Nachmani, E., Shaharbany, T., Wolf, L.: Segdiff: Image segmentation
  with diffusion probabilistic models. arXiv preprint arXiv:2112.00390  (2021)

\bibitem{baranchuk2021label}
Baranchuk, D., Rubachev, I., Voynov, A., Khrulkov, V., Babenko, A.:
  Label-efficient semantic segmentation with diffusion models. arXiv preprint
  arXiv:2112.03126  (2021)

\bibitem{chen2019automatic}
Chen, G., Xiang, D., Zhang, B., Tian, H., Yang, X., Shi, F., Zhu, W., Tian, B.,
  Chen, X.: Automatic pathological lung segmentation in low-dose ct image using
  eigenspace sparse shape composition. IEEE transactions on medical imaging
  \textbf{38}(7),  1736--1749 (2019)

\bibitem{choi2021ilvr}
Choi, J., Kim, S., Jeong, Y., Gwon, Y., Yoon, S.: Ilvr: Conditioning method for
  denoising diffusion probabilistic models. arXiv preprint arXiv:2108.02938
  (2021)

\bibitem{dhariwal2021diffusion}
Dhariwal, P., Nichol, A.: Diffusion models beat gans on image synthesis.
  Advances in Neural Information Processing Systems  \textbf{34} (2021)

\bibitem{gal2016dropout}
Gal, Y., Ghahramani, Z.: Dropout as a bayesian approximation: Representing
  model uncertainty in deep learning. In: international conference on machine
  learning. pp. 1050--1059. PMLR (2016)

\bibitem{guan2018said}
Guan, M., Gulshan, V., Dai, A., Hinton, G.: Who said what: Modeling individual
  labelers improves classification. In: Proceedings of the AAAI conference on
  artificial intelligence. vol.~32 (2018)

\bibitem{he2016deep}
He, K., Zhang, X., Ren, S., Sun, J.: Deep residual learning for image
  recognition. In: Proceedings of the IEEE conference on computer vision and
  pattern recognition. pp. 770--778 (2016)

\bibitem{ho2020denoising}
Ho, J., Jain, A., Abbeel, P.: Denoising diffusion probabilistic models.
  Advances in Neural Information Processing Systems  \textbf{33},  6840--6851
  (2020)

\bibitem{ho2022cascaded}
Ho, J., Saharia, C., Chan, W., Fleet, D.J., Norouzi, M., Salimans, T.: Cascaded
  diffusion models for high fidelity image generation. Journal of Machine
  Learning Research  \textbf{23}(47),  1--33 (2022)

\bibitem{huang2021variational}
Huang, C.W., Lim, J.H., Courville, A.C.: A variational perspective on
  diffusion-based generative models and score matching. Advances in Neural
  Information Processing Systems  \textbf{34} (2021)

\bibitem{jensen2019improving}
Jensen, M.H., J{\o}rgensen, D.R., Jalaboi, R., Hansen, M.E., Olsen, M.A.:
  Improving uncertainty estimation in convolutional neural networks using
  inter-rater agreement. In: Medical Image Computing and Computer Assisted
  Intervention--MICCAI 2019: 22nd International Conference, Shenzhen, China,
  October 13--17, 2019, Proceedings, Part IV 22. pp. 540--548. Springer (2019)

\bibitem{ji2021learning}
Ji, W., Yu, S., Wu, J., Ma, K., Bian, C., Bi, Q., Li, J., Liu, H., Cheng, L.,
  Zheng, Y.: Learning calibrated medical image segmentation via multi-rater
  agreement modeling. In: Proceedings of the IEEE/CVF Conference on Computer
  Vision and Pattern Recognition. pp. 12341--12351 (2021)

\bibitem{kingma2021variational}
Kingma, D.P., Salimans, T., Poole, B., Ho, J.: Variational diffusion models.
  arXiv preprint arXiv:2107.00630  (2021)

\bibitem{kohl2018probabilistic}
Kohl, S., Romera-Paredes, B., Meyer, C., De~Fauw, J., Ledsam, J.R., Maier-Hein,
  K., Eslami, S., Jimenez~Rezende, D., Ronneberger, O.: A probabilistic u-net
  for segmentation of ambiguous images. Advances in neural information
  processing systems  \textbf{31} (2018)

\bibitem{li2022srdiff}
Li, H., Yang, Y., Chang, M., Chen, S., Feng, H., Xu, Z., Li, Q., Chen, Y.:
  Srdiff: Single image super-resolution with diffusion probabilistic models.
  Neurocomputing  (2022)

\bibitem{long2015fully}
Long, J., Shelhamer, E., Darrell, T.: Fully convolutional networks for semantic
  segmentation. In: Proceedings of the IEEE conference on computer vision and
  pattern recognition. pp. 3431--3440 (2015)

\bibitem{loshchilov2017decoupled}
Loshchilov, I., Hutter, F.: Decoupled weight decay regularization. arXiv
  preprint arXiv:1711.05101  (2017)

\bibitem{nichol2021improved}
Nichol, A.Q., Dhariwal, P.: Improved denoising diffusion probabilistic models.
  In: International Conference on Machine Learning. pp. 8162--8171. PMLR (2021)

\bibitem{rahman2023ambiguous}
Rahman, A., Valanarasu, J.M.J., Hacihaliloglu, I., Patel, V.M.: Ambiguous
  medical image segmentation using diffusion models. In: Proceedings of the
  IEEE/CVF Conference on Computer Vision and Pattern Recognition. pp.
  11536--11546 (2023)

\bibitem{saharia2021image}
Saharia, C., Ho, J., Chan, W., Salimans, T., Fleet, D.J., Norouzi, M.: Image
  super-resolution via iterative refinement. arXiv preprint arXiv:2104.07636
  (2021)

\bibitem{san2021noise}
San-Roman, R., Nachmani, E., Wolf, L.: Noise estimation for generative
  diffusion models. arXiv preprint arXiv:2104.02600  (2021)

\bibitem{sohl2015deep}
Sohl-Dickstein, J., Weiss, E., Maheswaranathan, N., Ganguli, S.: Deep
  unsupervised learning using nonequilibrium thermodynamics. In: International
  Conference on Machine Learning. pp. 2256--2265. PMLR (2015)

\bibitem{wang2018esrgan}
Wang, X., Yu, K., Wu, S., Gu, J., Liu, Y., Dong, C., Qiao, Y., Change~Loy, C.:
  Esrgan: Enhanced super-resolution generative adversarial networks. In:
  Proceedings of the European conference on computer vision (ECCV) workshops.
  pp.~0--0 (2018)

\bibitem{wang2018deep}
Wang, Y., Deng, Z., Hu, X., Zhu, L., Yang, X., Xu, X., Heng, P.A., Ni, D.: Deep
  attentional features for prostate segmentation in ultrasound. In: Medical
  Image Computing and Computer Assisted Intervention--MICCAI 2018: 21st
  International Conference, Granada, Spain, September 16-20, 2018, Proceedings,
  Part IV 11. pp. 523--530. Springer (2018)

\bibitem{wolleb2021diffusion}
Wolleb, J., Sandkühler, R., Bieder, F., Valmaggia, P., Cattin, P.C.: Diffusion
  models for implicit image segmentation ensembles (2021)

\bibitem{zhou2017fixed}
Zhou, Y., Xie, L., Shen, W., Wang, Y., Fishman, E.K., Yuille, A.L.: A
  fixed-point model for pancreas segmentation in abdominal ct scans. In:
  Medical Image Computing and Computer Assisted Intervention- MICCAI 2017: 20th
  International Conference, Quebec City, QC, Canada, September 11-13, 2017,
  Proceedings, Part I. pp. 693--701. Springer (2017)

\end{thebibliography}
%


\end{document}